\documentclass[preprintnumbers,eqsecnum,prd,twocolumn,showpacs,showkeys,nofootinbib]{revtex4}  

\usepackage{graphicx}
\usepackage{latexsym}
\usepackage{amsmath}
\usepackage{amssymb}

\def\beq{\begin{equation}}
\def\eeq{\end{equation}}

\begin{document}

\title{Scalar field self-force effects on a particle orbiting a Reissner-Nordstr\"om black hole}

\author{Donato Bini}
  \affiliation{
Istituto per le Applicazioni del Calcolo ``M. Picone,'' CNR, I-00185 Rome, Italy
}
\email{donato.bini@gmail.com}

\author{Gabriel Carvalho}
  \affiliation{CAPES Foundation, Ministry of Education of Brazil, Bras\'ilia, Brazil and\\
Sapienza Universit\`a di Roma - Dipartimento di Fisica, P.le Aldo Moro 5, 00185 Rome,  Italy
} 
\email{gabriel.carvalho@icranet.org}

\author{Andrea Geralico}
\affiliation{
Astrophysical Observatory of Torino, INAF, 
I-10025 Pino Torinese (TO), Italy
}
\email{andrea.geralico@gmail.com}

\date{\today}

\begin{abstract}
Scalar field self-force effects on a scalar charge orbiting a Reissner-Nordstr\"om black hole are investigated. 
The scalar wave equation is solved analytically in a post-Newtonian framework, and the solution is used to compute the self-field as well as the components of the self-force at the particle's location up to 7.5 post-Newtonian order. The energy fluxes radiated to infinity and down the hole are also evaluated.
Comparison with previous numerical results in the Schwarzschild case shows a good agreement in both strong-field and weak-field regimes.
\end{abstract}


\maketitle

\section{Introduction}

Scalar self-force (SSF) effects arise when a scalar charge, moving along a given orbit in a curved spacetime, interacts with its own gravitational field, i.e., its self-field. The associated scalar field satisfies a d'Alembert-like equation with source term singular at the particle's position, mimicking the more interesting situation of gravitational perturbations induced by a small mass moving in a gravitational background modified by its own presence. The interaction of the particle with its own gravitational field in this case gives rise to a gravitational self-force (GSF) (see, e.g., Ref. \cite{Barack:2009ux} and references therein).
It is a matter of fact that the latter problem is physically more interesting than the first one. However, the study of the first problem is easier than the second, even if the approaches as well as the computational techniques used in both cases are similar.
This explains why in the literature the SSF problem has been considered as a preliminary study to the GSF one, scouting/solving all technical difficulties also affecting the more general gravitational perturbation problem.
The existing literature on this topic is very rich. Indeed, besides the various pioneering works developing the fundamental formalism for self-force calculations in a
curved spacetime \cite{Mino:1996nk,Quinn:1996am,Quinn:2000wa,Barack:2001gx,Barack:2001ph,Detweiler:2002mi,Poisson:2003nc}, a number of interesting papers has been produced over the years, aiming at understanding self-force effects in black hole spacetimes, mostly Schwarzschild and Kerr~\cite{Burko:1999zy,Barack:2000eh,Burko:2000xx,Wiseman:2000rm,Nakano:2001kw,Burko:2001kr,Barack:2002mha,Detweiler:2002gi,Nakano:2003he,Hikida:2003pi,DiazRivera:2004ik,Hikida:2004hs,Ottewill:2007mz,Haas:2007kz,Damour:2009sm,Barack:2010tm,Warburton:2010eq,Warburton:2011hp,Wardell:2011gb,Ottewill:2012aj,Vega:2013wxa,Isoyama:2014mja,Warburton:2014bya,vandeMeent:2016pee}.

The present paper concerns scalar field self-force effects in a Reissner-Nordstr\"om (RN) spacetime on a scalar charge moving along a spatially circular equatorial orbit around a spherically symmetric charged black hole. The interaction between the particle and the background field is thus of the gravitational type only, the particle carrying no electromagnetic charge. We are interested in studying the coupling between the scalar charge of the associated field with the mass and the electromagnetic charge of the non-vacuum background, which was never explored before.
Switching off the black hole charge one ends up with the corresponding SSF problem in the vacuum Schwarzschild spacetime \cite{Nakano:2001kw,Detweiler:2002gi,DiazRivera:2004ik,Hikida:2004hs}.

The main technical difficulties associated with self-force calculations are related to the regularization procedure of the scalar field and its derivatives, allowing to extract the correct, physically meaningful self force components. 
We use the standard post-Newtonian (PN) expansion method to compute the self-field decomposed into spherical harmonics and frequency modes, and regularize it at the particle's position mode by mode by subtracting the diverging large-$l$ limit.     
We analytically compute the regularized self-field as well as the components of the self-force up to 7.5PN order and compare our results with previous numerical studies in the Schwarzschild case \cite{DiazRivera:2004ik,Warburton:2010eq}, obtaining a good agreement also in the strong-field regime. 
Finally, we complete our analysis by providing explicit expressions for the scalar radiation both at infinity and on the outer horizon.
Again, the comparison with previous numerical results in the Schwarzschild case \cite{Warburton:2010eq} shows a good agreement.

\section{Scalar charge in a Reissner-Nordstr\"om background}

Let us consider a Reissner-Nordstr\"om spacetime with line element written in standard Schwarzschild-like coordinates $(t,r,\theta,\phi)$ as
\beq
ds^2 =-\frac{\Delta}{r^2}dt^2 + \frac{r^2}{\Delta}dr^2 + r^2(d\theta^2+\sin^2\theta d\phi^2)\,,
\eeq
where $\Delta= r^2-2Mr+Q^2$.
The condition $\Delta=0$ defines the two horizons at radii $r_\pm=M\pm \sqrt{M^2-Q^2}\equiv M(1 \pm  \kappa)$, with $\kappa=\sqrt{1-Q^2/M^2}$.
The ``extreme'' case corresponds to $|Q|=M$ (or $\kappa=0$), the two horizons coalescing into one. We find it convenient to introduce also the notation
$w=1-\kappa^2=(Q/M)^2$, such that $w=0$ corresponds to the Schwarzschild limit, while $w=1$ to the extreme RN case.

Let $\psi$ be a (real, minimally coupled) scalar field associated with a scalar charge $q$ moving along a circular equatorial timelike geodesic with 4-velocity $U=\Gamma(\partial_t + \Omega \partial_{\phi})$ and parametric equations  $x^\mu =z^{\mu}(\tau)$
\beq
\label{worldline}
t=\Gamma \tau\,,\qquad r=r_0\,, \qquad \theta=\frac{\pi}{2}\,,\quad
\phi = \Gamma\Omega \tau = \Omega t\,,
\eeq
where $\tau$ denotes the proper time, and the normalization factor $\Gamma$ and the angular velocity $\Omega$ are conveniently written in terms of the inverse dimensionless radial distance $u=M/r$ as 
\begin{eqnarray}
\label{relations}
\Gamma  &=& \frac{1}{\sqrt{1-3u +(1-\kappa^2)u^2}}\,,\nonumber\\
M\Omega &=& u^{3/2}\, \sqrt{1-(1-\kappa^2) u}\,,
\end{eqnarray}
respectively.

Assuming that the particle's field can be treated as a small perturbation on the fixed RN background implies that it obeys the massless Klein-Gordon equation
\beq
\label{fundam_eq}
\square \psi = -4\pi\varrho\,,
\eeq
where 
\beq
\square \psi = \frac{1}{\sqrt{-g}} \partial_{\mu}(\sqrt{-g}g^{\mu \nu}\partial_{\nu}\psi)
\eeq
is the D'Alembertian (with $g$ denoting the determinant of the metric) and 
\begin{eqnarray}
\varrho(x^\mu) &=& q\int{ (-g)^{-1/2} \delta^4(x^\mu - z^{\mu}(\tau))}d\tau\nonumber\\
&=& \frac{q}{r_0^2 \Gamma}\delta(r-r_0) \delta\left(\theta - \frac{\pi}{2}\right) \delta(\phi - \Omega t)\,,
\end{eqnarray}
the charge density of the scalar particle with support only along the particle's world line \eqref{worldline}.
Decomposing into spherical harmonics then gives 
\beq
\label{rhodef}
\varrho=\frac{1}{4\pi r_0}\delta(r-r_0)\sum_{lm}q_{lm}e^{-im\Omega t} Y_{lm}(\theta,\phi)\,,
\eeq
where 
\beq
q_{lm} = \frac{4\pi q}{r_0 \Gamma} Y_{lm}^* (\pi/2,0)\,,
\eeq
and similarly for the scalar field $\psi$, whose dependence on temporal, radial and angular variables can be separated as
\beq
\label{psidef}
\psi(t,r,\theta,\phi) =\frac1{2\pi}\int\sum_{lm} \psi_{lm\omega}(r)e^{-i\omega t}Y_{lm}(\theta, \phi)\,d\omega\,.
\eeq
The wave equation \eqref{fundam_eq} thus reduces to the following equation for the radial part
\beq
\label{radialeq}
{\mathcal L}_{(r)}(\psi_{lm\omega}(r))= S_{lm\omega} \delta(r-r_0)\,,
\eeq
with 
\begin{eqnarray}
{\mathcal L}_{(r)}(\psi_{lm\omega}(r))&\equiv&
\frac{d^2}{dr^2}\psi_{lm\omega}(r) + \frac{2(r-M)}{\Delta} \frac{d}{dr}\psi_{lm\omega}(r) \nonumber\\
&+ &  \left[\frac{\omega^2 r^4}{\Delta^2} - \frac{l(l+1)}{\Delta} \right]\psi_{lm\omega}(r)\,,
\end{eqnarray}
whereas
\beq
S_{lm\omega} = -2\pi \frac{r_0 }{\Delta_0}q_{lm}\delta(\omega-m\Omega)\,,
\eeq
with $\Delta_0=\Delta(r_0)$, comes from taking the Fourier-transform of the charge density \eqref{rhodef}.

\section{Computation of the scalar field along the world line}

The radial part of the scalar field is computed by using the Green function method as
\begin{eqnarray}
\psi_{lm\omega}(r) &=& \int G_{lm\omega}(r,r')\Delta(r')S_{lm\omega}\delta(r' - r_0)dr'\nonumber\\
&=& G_{lm\omega}(r,r_0)\Delta_0S_{lm\omega}\,,
\end{eqnarray}
where the Green function $G_{lm\omega}(r,r')$ satisfies the equation ${\mathcal L}_{(r)}(G_{lm\omega}(r,r'))=\Delta^{-1}(r')\delta(r-r')$.
It reads as
\begin{eqnarray}
G_{lm\omega}(r,r')&=&\frac{1}{W_{lm\omega}}\left[R_{\rm in}^{lm\omega}(r)R_{\rm up}^{lm\omega}(r')H(r'-r)\right. \nonumber\\
&& \left. + R_{\rm in}^{lm\omega}(r')R_{\rm up}^{lm\omega}(r)H(r-r') \right]\,,
\end{eqnarray}
where $H(x)$ denotes the Heaviside step function, $R_{\rm in}^{lm\omega}(r)$ and $R_{\rm up}^{lm\omega}(r)$ are two independent homogeneous solutions of the radial wave equation having the correct behavior at the outer horizon and at infinity, respectively, and  
\beq
W_{lm\omega}=\Delta(r)\left[R_{\rm in}^{lm\omega}(r)R'{}_{\rm up}^{lm\omega}(r)-R'{}_{\rm in}^{lm\omega}(r)R_{\rm up}^{lm\omega}(r)\right]
\eeq
is the associated (constant) Wronskian.
Substituting then into Eq. \eqref{psidef} gives
\beq
\psi(x^\mu)=
 -\sum_{lm} G_{lm\omega}(r,r_0)\Big\vert_{\omega=m\Omega} r_0 q_{lm}e^{-im\Omega t}Y_{lm}(\theta,\phi)\,,
\eeq
which, once evaluated along the particle world line \eqref{worldline}, becomes
\beq
\label{psi0def}
\psi_0=-\frac{4\pi q}{\Gamma} \sum_{lm}G_{lm\omega}(r_0,r_0)\Big\vert_{\omega=m\Omega}|Y_{lm}(\pi/2,0)|^2\,,
\eeq
only depending on $r_0$.
The above expression for $\psi_0$ actually requires taking the limit $r\to r_0^\pm$ properly, and must be suitably regularized in order to remove its
singular part, because the field has a divergent behavior there.

In order to compute the Green function we have first to solve the homogeneous radial wave equation \eqref{radialeq} up to a certain PN order to obtain the in and up  solutions, which are of the form
\begin{eqnarray}
\label{PNsols}
R_{\rm in(PN)}^{lm\omega}(r) &=& r^l [1 + A_2^{lm\omega}(r)\eta^2 + A_4^{lm\omega}(r)\eta^4 \nonumber\\
&& + A_6^{lm\omega}(r)\eta^6 + A_8^{lm\omega}(r)\eta^8 + \ldots]\,,\nonumber\\
R_{\rm up(PN)}^{lm\omega}(r) &=& R_{\rm in(PN)}^{-l-1m\omega}(r)\,. 
\end{eqnarray}
However, these solutions do not automatically fulfill the correct boundary conditions. 
A consequence of this fact is the presence of diverging terms in the coefficients $A_i$ for certain values of $l$. 
Therefore, high-order PN solutions usually require using a technique first introduced by Mano, Suzuki and Takasugi (MST) \cite{Mano:1996mf,Mano:1996vt}.
We will show some detail in Appendix A.

Turning then to Eq. \eqref{psi0def}, the sum over $m$ is straightforwardly computed by using standard formulas.
Before summing over $l$, instead, one has to remove the divergent term for large $l$, i.e.,
\beq
\psi_0^{\rm reg}=\sum_l(\psi_0^l-B)\,.
\eeq
The subtraction term turns out to be (in units of $q$)
\begin{eqnarray}
B&=&u-\frac{1}{4}u^2
+\left(\frac{9}{64}-\frac{3}{4}\kappa^2\right)u^3\nonumber\\
&&
+\left(-\frac{73}{32}\kappa^2+\frac{199}{256}\right)u^4\nonumber\\
&&
+\left(\frac{39625}{16384}-\frac{39}{64}\kappa^4-\frac{1425}{256}\kappa^2\right)u^5\nonumber\\
&&
+\left(-\frac{907}{256}\kappa^4-\frac{52585}{4096}\kappa^2+\frac{451007}{65536}\right)u^6\nonumber\\
&&
+\left(-\frac{1926415}{65536}\kappa^2-\frac{109317}{8192}\kappa^4\right.\nonumber\\
&&\left.
+\frac{20043121}{1048576}-\frac{171}{256}\kappa^6\right)u^7
+O(u^8)\,.
\end{eqnarray}
This can be shown to be the Taylor expansion of  
\beq
B_{\rm analytic}=\frac{u}{\sqrt{1-3u}}\frac{\sqrt{1-\sigma}}{\Gamma}\frac{2}{\pi}{\rm EllipticK}(\sigma)\,,
\eeq
where
\beq
\sigma=\frac{u[1+u(1-\kappa^2)]}{1-2u+u^2(1-\kappa^2)}\,.
\eeq

It is useful to introduce the dimensionless angular velocity variable 
\beq
y=(M\Omega)^{2/3}=u(1+wu)^{1/3}\,,
\eeq
as from Eq. \eqref{relations}, with inverse relation
\begin{eqnarray}
\label{uvsy}
u &=& y-\frac{1}{3}wy^2+\frac{1}{3}w^2y^3-\frac{35}{81}w^3y^4+\frac{154}{243}w^4y^5\nonumber\\
&& -w^5y^6+\frac{10868}{6561}w^6y^7+O(y^8)\,,
\end{eqnarray}
where we recall $w=1-\kappa^2$.

By applying the MST technique \cite{Mano:1996mf,Mano:1996vt} to the multipoles up to $l=4$ (included), we get the following final result for the regularized field valid up to the 7.5 PN order
\begin{widetext}
\begin{eqnarray}
\label{psi_0_finale}
\psi_0^{\rm reg}&=&-y^3
+\left[
\frac{35}{18}+\left(-\frac{7}{32}+\frac{w}{32}\right)\pi^2-\frac{4}{3}\gamma-\frac{4}{3}\ln(2)-\frac{2}{3}\ln(y)
\right]y^4\nonumber\\
&&
+\left[
\frac{1141}{360}-\frac{35}{54}w
+\left(\frac{29}{512}+\frac{97}{1536}w-\frac{w^2}{96}\right)\pi^2
+\left(\frac{2}{3}-\frac{8}{9}w\right)\gamma
+\left(-\frac{18}{5}-\frac{8}{9}w\right)\ln(2)
+\left(\frac{1}{3}-\frac{4}{9}w\right)\ln(y)
\right]y^5\nonumber\\
&&
+\left(-\frac{38}{45}+\frac{8}{45}w\right)\pi y^{11/2}
\nonumber\\
&&
+\left[
-\frac{23741}{1680}+\frac{4607}{540}w+\frac{23}{54}w^2
+\left(-\frac{279}{1024}-\frac{397}{1536}w-\frac{11}{512}w^2+\frac{1}{96}w^3\right)\pi^2
+\left(\frac{77}{6}-\frac{46}{9}w+\frac{4}{9}w^2\right)\gamma\right.\nonumber\\
&&\left.
+\left(\frac{1627}{42}-\frac{54}{5}w+\frac{4}{9}w^2\right)\ln(2)
-\frac{729}{70}\ln(3)
+\left(\frac{77}{12}-\frac{23}{9}w+\frac{2}{9}w^2\right)\ln(y)
\right]y^6\nonumber\\
&&
+\left(-\frac{3}{35}-\frac{2696}{4725}w+\frac{16}{135}w^2\right)\pi y^{13/2}
\nonumber\\
&&
+\left\{
-\frac{1515589307}{27216000}+\frac{3098381}{378000}w-\frac{3497}{3240}w^2-\frac{793}{1458}w^3\right.\nonumber\\
&&\left.
+\left(-\frac{58}{45}+\frac{8}{45}w\right)(1-w)^{3/2}
-\frac23 (2-w) (1-w)\ln(1-w)\right.\nonumber\\
&&
+\left(-\frac{6059603}{983040}+\frac{1892003}{983040}w+\frac{2287}{9216}w^2+\frac{871}{20736}w^3-\frac{35}{2592}w^4\right)\pi^2
+\left(\frac{76585}{262144}-\frac{14281}{131072}w+\frac{2665}{262144}w^2\right)\pi^4\nonumber\\
&&
+\left[
-\frac{5321}{900}+\frac{4312}{675}w-\frac{4}{27}w^2-\frac{112}{243}w^3
+\left(\frac{152}{45}-\frac{32}{45}w\right)\gamma
+\left(\frac{304}{45}-\frac{64}{45}w\right)\ln(2)
+\left(\frac{152}{45}-\frac{32}{45}w\right)\ln(y)
\right]\gamma\nonumber\\
&&
+\left[
-\frac{1786621}{18900}+\frac{149404}{4725}w-\frac{8}{15}w^2-\frac{112}{243}w^3
+\left(\frac{152}{45}-\frac{32}{45}w\right)\ln(2)
+\left(\frac{152}{45}-\frac{32}{45}w\right)\ln(y)
\right]\ln(2)\nonumber\\
&&\left.
+\left(\frac{12393}{140}-\frac{729}{35}w\right)\ln(3)
-\frac{16}{3}\zeta(3)
+\left[
-\frac{10121}{1800}+\frac{4856}{675}w-\frac{38}{27}w^2-\frac{56}{243}w^3
+\left(\frac{38}{45}-\frac{8}{45}w\right)\ln(y)
\right]\ln(y)
\right\}y^7\nonumber\\
&&
+\left(\frac{35633}{3780}-\frac{192541}{33075}w+\frac{5062}{4725}w^2-\frac{8}{135}w^3\right)\pi y^{15/2}
+O(y^8)\,.
\end{eqnarray}

In the Schwarzschild case (i.e., in the limit $w\to 0$) it reduces to
\begin{eqnarray}
\label{psi0schwy}
\psi_0^{\rm reg,\,schw}&=&-y^3
+\left(
\frac{35}{18}-\frac{7}{32}\pi^2-\frac{4}{3}\gamma-\frac{4}{3}\ln(2)-\frac{2}{3}\ln(y)
\right)y^4
+\left(
\frac{1141}{360}
+\frac{29}{512}\pi^2
+\frac{2}{3}\gamma
-\frac{18}{5}\ln(2)
+\frac{1}{3}\ln(y)
\right)y^5\nonumber\\
&&
-\frac{38}{45}\pi y^{11/2}
\nonumber\\
&&
+\left(
-\frac{23741}{1680}
-\frac{279}{1024}\pi^2
+\frac{77}{6}\gamma
+\frac{1627}{42}\ln(2)
-\frac{729}{70}\ln(3)
+\frac{77}{12}\ln(y)
\right)y^6\nonumber\\
&&
-\frac{3}{35}\pi y^{13/2}
\nonumber\\
&&
+\left[
-\frac{1515589307}{27216000}
-\frac{6059603}{983040}\pi^2
+\frac{76585}{262144}\pi^4
+\left(
-\frac{5321}{900}
+\frac{152}{45}\gamma
+\frac{304}{45}\ln(2)
+\frac{152}{45}\ln(y)
\right)\gamma\right.\nonumber\\
&&\left.
+\left(
-\frac{1786621}{18900}
+\frac{152}{45}\ln(2)
+\frac{152}{45}\ln(y)
\right)\ln(2)
+\frac{12393}{140}\ln(3)
-\frac{16}{3}\zeta(3)
+\left(
-\frac{10121}{1800}
+\frac{38}{45}\ln(y)
\right)\ln(y)
\right]y^7\nonumber\\
&&
+\frac{35633}{3780}\pi y^{15/2}
+O(y^8)\,,
\end{eqnarray}
which was never shown before in the literature \footnote{
The first terms of this expansion (up to $O(y^5)$ included) agree with unpublished results by Bini and Damour \cite{BD_unpub}.
}.
Scalar self force effects on a Schwarzschild background was numerically studied in Ref. \cite{DiazRivera:2004ik}. We show in Table \ref{tab:1} and in Fig. \ref{fig:1} the comparison between our analytical results and the numerical values mentioned above. The agreement is excellent (i.e., of the order $\sim 10^{-14}$) in the weak-field region, as expected, and also good enough (i.e., of the order $\sim 10^{-4}$) in the strong-field region.

In order to study the transcendental structure of the various PN orders, it is useful to replace ordinary logarithms by \lq\lq eulerlogs,'' i.e.,
\beq
{\rm eulerlog}_m(x)=\gamma+\ln(2)+\frac12\ln(y)+\ln(m)\,, \qquad 
m=1,2,3,\ldots\,,
\eeq
first introduced in Ref. \cite{Damour:2008gu}, so to absorb also the Euler $\gamma$ constant.
We obtain
\begin{eqnarray}
\psi_0^{\rm reg,\,schw}&=&-y^3
+\left(
\frac{35}{18}-\frac{7}{32}\pi^2-\frac{4}{3}{\rm eulerlog}_1(y)
\right)y^4
+\left(
\frac{1141}{360}
+\frac{29}{512}\pi^2
+\frac{74}{15}{\rm eulerlog}_1(y)
-\frac{64}{15}{\rm eulerlog}_2(y)
\right)y^5\nonumber\\
&&
-\frac{38}{45}\pi y^{11/2}
\nonumber\\
&&
+\left(
-\frac{23741}{1680}
-\frac{279}{1024}\pi^2
-\frac{93}{35}{\rm eulerlog}_1(y)
+\frac{544}{21}{\rm eulerlog}_2(y)
-\frac{729}{70}{\rm eulerlog}_3(y)
\right)y^6\nonumber\\
&&
-\frac{3}{35}\pi y^{13/2}
\nonumber\\
&&
+\left[
-\frac{1515589307}{27216000}
-\frac{6059603}{983040}\pi^2
+\frac{76585}{262144}\pi^4
+\frac{16}{3}\gamma
-\frac{16}{3}\zeta(3)
-\frac{77879}{4725}{\rm eulerlog}_1(y)
-\frac{78704}{945}{\rm eulerlog}_2(y)
\right.\nonumber\\
&&\left.
+\frac{12393}{140}{\rm eulerlog}_3(y)
+\frac{152}{45}{\rm eulerlog}_1^2(y)
\right]y^7\nonumber\\
&&
+\frac{35633}{3780}\pi y^{15/2}
+O(y^8)\,.
\end{eqnarray}
Unfortunately, this replacement is not enough to completely remove the Euler $\gamma$ term at the order $O(y^7)$, meaning that the transcendental structure is more involved than simple eulerlogs.

It is also interesting to study the behavior of this scalar field at the light-ring $y=1/3$.
A simple numerical fit 
\beq
\psi_0^{\rm reg,\,schw\,fit}=-\frac{y^3}{(1-3y)^2}
(1-7.84 y+47.36 y^2-8.65 y^3+81.77 y^3\ln(y))
\eeq
(with a maximal residual of about $2.4\times10^{-4}$) suggests a blow up of the form $(1-3y)^{-2}$.
However, this is an indication only, and a more conclusive statement requires strong field numerical data still currently unavailable.

Finally, in the extreme RN case (i.e., in the limit $w\to 1$) we have
\begin{eqnarray}
\psi_0^{\rm reg,\,extr}&=&-y^3
+\left(
\frac{35}{18}-\frac{3}{16}\pi^2-\frac{4}{3}\gamma-\frac{4}{3}\ln(2)-\frac{2}{3}\ln(y)
\right)y^4
+\left(
\frac{2723}{1080}
+\frac{7}{64}\pi^2
-\frac{2}{9}\gamma
-\frac{202}{45}\ln(2)
-\frac{1}{9}\ln(y)
\right)y^5\nonumber\\
&&
-\frac{2}{3}\pi y^{11/2}
\nonumber\\
&&
+\left(
-\frac{78233}{15120}
-\frac{555}{1024}\pi^2
+\frac{49}{6}\gamma
+\frac{17881}{630}\ln(2)
-\frac{729}{70}\ln(3)
+\frac{49}{12}\ln(y)
\right)y^6\nonumber\\
&&
-\frac{121}{225}\pi y^{13/2}
\nonumber\\
&&
+\left[
-\frac{160402001}{3265920}
-\frac{438259}{110592}\pi^2
+\frac{99}{512}\pi^4
+\left(
-\frac{647}{4860}
+\frac{8}{3}\gamma
+\frac{16}{3}\ln(2)
+\frac{8}{3}\ln(y)
\right)\gamma\right.\nonumber\\
&&\left.
+\left(
-\frac{2174033}{34020}
+\frac{8}{3}\ln(2)
+\frac{8}{3}\ln(y)
\right)\ln(2)
+\frac{9477}{140}\ln(3)
-\frac{16}{3}\zeta(3)
+\left(
-\frac{647}{9720}
+\frac{2}{3}\ln(y)
\right)\ln(y)
\right]y^7\nonumber\\
&&
+\frac{203629}{44100}\pi y^{15/2}
+O(y^8)\,.
\end{eqnarray}
[Note that the term with $\ln(1-w)$ in Eq. \eqref{psi_0_finale} is proportional to $(1-w)\ln(1-w)$, which vanishes in the limit $w\to 1$, so that the final expression is finite.]

  
\begin{table}
\centering
\caption{
Comparison between the analytical prediction \eqref{psi0schwy} for the regularized scalar field in the Schwarzschild case ($w=0$) and the numerical values taken from Table I of Ref. \cite{DiazRivera:2004ik}.
The difference $\Delta\psi_0^{\rm schw}=\psi_0^{\rm schw,\,num}-\psi_0^{\rm schw}$ is shown in the 3rd column.
}
\begin{ruledtabular}
\begin{tabular}{ccc}
$y$ & $\psi_0^{\rm schw}$ & $\Delta\psi_0^{\rm schw}$  \cr
\hline
 1/4 & $-0.02304519610$ & $-9.43\times10^{-4}$ \cr
 1/5  & $-0.01022371010$ & $-1.05\times10^{-5}$ \cr
 1/6  & $-0.005468782560$ & $1.40\times10^{-5}$ \cr
 1/7  & $-0.003282635718$ & $7.29\times10^{-6}$ \cr
 1/8  & $-0.002130877461$ & $3.37\times10^{-6}$ \cr
 1/10  & $-0.001050586634$ & $7.94\times10^{-7}$  \cr
 1/14  & $-3.701411742\times10^{-4}$ & $7.66\times10^{-8}$  \cr
 1/20   &$ -1.246786056\times10^{-4}$ & $5.81\times10^{-9}$  \cr
 1/30   & $-3.661740186\times10^{-5}$ & $3.02\times10^{-10}$  \cr
 1/50   & $-7.889525256\times10^{-6}$ & $7.26\times10^{-12}$  \cr
 1/70   & $-2.877222881\times10^{-6}$ & $8.81\times10^{-13}$  \cr
 1/100   &$ -9.884245218\times10^{-7}$ & $2.18\times10^{-14}$  \cr
 1/200   & $-1.239865750\times10^{-7}$ & $-2.50\times10^{-14}$  \cr
\end{tabular}
\end{ruledtabular}
\label{tab:1}
\end{table}


\begin{figure}
\begin{center}
\includegraphics[scale=0.4]{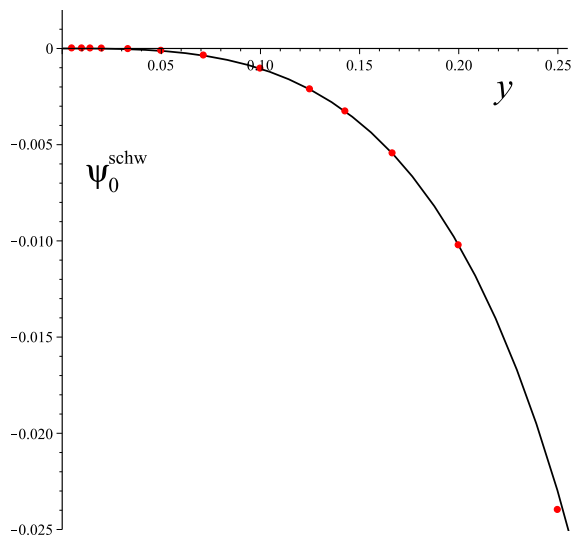} 
\end{center}
\caption{
The behavior of the regularized scalar field \eqref{psi0schwy} in the Schwarzschild case ($w=0$) as a function of $y$ is shown in comparison with existing numerical values. The data points are taken from Table I of Ref. \cite{DiazRivera:2004ik}.
}
\label{fig:1}
\end{figure}

\section{Scalar self-force}

The scalar self-force is given by
\begin{eqnarray}
F_\alpha(x^\mu)&=&q\nabla_\alpha\psi(x^\mu)\nonumber\\
&=&-q\nabla_\alpha\sum_{lm} G_{lm\omega}(r,r_0)\Big\vert_{\omega=m\Omega} r_0 q_{lm}e^{-im\Omega t}Y_{lm}(\theta,\phi)\,,
\end{eqnarray}
with nonvanishing components 
\begin{eqnarray}
F_{t\,(\pm)}^0&=&-\Omega F_{\phi\,(\pm)}^0
=i\Omega\frac{4\pi q^2}{\Gamma} \sum_{lm}m\,G_{lm\omega}(r,r_0)\Big\vert_{r=r_0^\pm,\,\omega=m\Omega}|Y_{lm}(\pi/2,0)|^2\,,\nonumber\\
F_{r\,(\pm)}^0&=&-\frac{4\pi q^2}{\Gamma} \sum_{lm}\partial_rG_{lm\omega}(r,r_0)\Big\vert_{r=r_0^\pm,\,\omega=m\Omega}|Y_{lm}(\pi/2,0)|^2\,,
\end{eqnarray}
once evaluated at the position of the scalar charge, i.e., in the limit $r\to r_0^\pm$.
After summing over $m$, the divergent behavior for large $l$ is removed by 
\beq
F_\alpha^{0\,\rm reg}=\sum_l\left[\frac12\left(F_{\alpha\,(+)}^{0l}+F_{\alpha\,(-)}^{0l}\right)-B_\alpha\right]\,,
\eeq
where $F_{\alpha\,(\pm)}^{0l}$ denote the limits $r\to r_0^\pm$ of each mode and the $l$-independent quantities $B_\alpha$ are suitable regularization parameters.
The subtraction term for the radial component turns out to be (in units of $q$)
\begin{eqnarray}
B_r&=&-\frac12y^2
+\left(-\frac18+\frac13w\right)y^3
+\left(-\frac{21}{128}+\frac12w-\frac{7}{18}w^2\right)y^4
+\left(-\frac{53}{512}+\frac{57}{64}w-\frac23w^2+\frac{44}{81}w^3\right)y^5\nonumber\\
&&
+\left(\frac{12607}{32768}+\frac{97}{96}w-\frac{331}{384}w^2+w^3-\frac56w^4\right)y^6\nonumber\\
&&
+\left(\frac{306759}{131072}-\frac{18433}{16384}w+\frac{4517}{2304}w^2+\frac{1055}{864}w^3-\frac{130}{81}w^4+\frac{988}{729}w^5\right)y^7
+O(y^8)\,,
\end{eqnarray}
whereas $B_t=0$.

The final result for the regularized temporal and radial components of the self force valid through 7.5PN order is (in units of $q$)
\begin{eqnarray}
F_t^{0\,\rm reg}&=&
\frac13y^4+\left(-\frac16+\frac29w\right)y^5+\frac23\pi y^{11/2}+\left(-\frac{77}{24}+\frac{23}{18}w-\frac19w^2\right)y^6+\left(\frac95+\frac49w\right)\pi y^{13/2}\nonumber\\
&&
+\left[
\frac{7721}{3600}-\frac{1753}{675}w+\frac{10}{27}w^2+\frac{28}{243}w^3+\frac23(1-w)^{3/2}+\frac49\pi^2
+\left(-\frac{76}{45}+\frac{16}{45}w\right)\gamma
+\left(-\frac{76}{45}+\frac{16}{45}w\right)\ln(2)\right.\nonumber\\
&&\left.
+\left(-\frac{38}{45}+\frac{8}{45}w\right)\ln(y)
\right]y^7\nonumber\\
&&
+\left(-\frac{3761}{420}+\frac{27}{5}w-\frac29w^2\right)\pi y^{15/2}
+O(y^8)
\,,\nonumber\\
F_r^{0\,\rm reg}&=&
\left[-\frac29+\left(\frac{7}{64}-\frac{1}{64}w\right)\pi^2-\frac43\gamma-\frac43\ln(2)-\frac23\ln(y)\right]y^5
\nonumber\\
&&
+\left[
\frac{604}{45}-\frac{41}{27}w+\left(\frac{29}{1024}-\frac{239}{3072}w+\frac{1}{96}w^2\right)\pi^2-\left(\frac{14}{3}+\frac49w\right)\gamma-\left(\frac{66}{5}+\frac49w\right)\ln(2)-\left(\frac73+\frac29w\right)\ln(y)
\right]y^6\nonumber\\
&&
+\left(-\frac{38}{45}+\frac{8}{45}w\right)\pi y^{13/2}\nonumber\\
&&
+\left[
\frac{1511}{140}+\frac{473}{90}w+\frac{103}{81}w^2
+\left(\frac{1335}{2048}-\frac{1}{16}w+\frac{151}{2304}w^2-\frac{7}{576}w^3\right)\pi^2
+\left(\frac{31}{2}-\frac{28}{3}w+\frac{8}{27}w^2\right)\gamma\right.\nonumber\\
&&\left.
+\left(\frac{857}{14}-\frac{268}{15}w+\frac{8}{27}w^2\right)\ln(2)-\frac{2187}{70}\ln(3)
+\left(\frac{31}{4}-\frac{14}{3}w+\frac{4}{27}w^2\right)\ln(y)
\right]y^7\nonumber\\
&&
+\left(-\frac{139}{35}+\frac{2378}{4725}w+\frac{8}{135}w^2\right)\pi y^{15/2}
+O(y^8)
\,,
\end{eqnarray}
respectively.

In the Schwarzschild limit ($w\to0$) we have
\begin{eqnarray}
\label{SSFschw}
F_t^{0\,\rm reg,\,schw}&=&
\frac13y^4-\frac16y^5+\frac23\pi y^{11/2}-\frac{77}{24}y^6+\frac95\pi y^{13/2}\nonumber\\
&&
+\left[
\frac{10121}{3600}+\frac49\pi^2
-\frac{76}{45}\gamma
-\frac{76}{45}\ln(2)
-\frac{38}{45}\ln(y)
\right]y^7\nonumber\\
&&
-\frac{3761}{420}\pi y^{15/2}
+O(y^8)
\,,\nonumber\\
F_r^{0\,\rm reg,\,schw}&=&
\left[-\frac29+\frac{7}{64}\pi^2-\frac43\gamma-\frac43\ln(2)-\frac23\ln(y)\right]y^5
\nonumber\\
&&
+\left[
\frac{604}{45}+\frac{29}{1024}\pi^2-\frac{14}{3}\gamma-\frac{66}{5}\ln(2)-\frac73\ln(y)
\right]y^6\nonumber\\
&&
-\frac{38}{45}\pi y^{13/2}\nonumber\\
&&
+\left[
\frac{1511}{140}
+\frac{1335}{2048}\pi^2
+\frac{31}{2}\gamma
+\frac{857}{14}\ln(2)-\frac{2187}{70}\ln(3)
+\frac{31}{4}\ln(y)
\right]y^7\nonumber\\
&&
-\frac{139}{35}\pi y^{15/2}
+O(y^8)
\,.
\end{eqnarray}
The leading 3PN and 4PN terms of the previous expressions agree with those of Ref. \cite{Hikida:2004hs}.
Furthermore, the comparison with available numerical results of Ref. \cite{DiazRivera:2004ik,Warburton:2010eq} shows again a very good agreement (see Table \ref{tab:x} and Fig. \ref{fig:2}, where we refer to the most recent work \cite{Warburton:2010eq}).

  
\begin{table}
\centering
\caption{
Comparison between the analytical expressions \eqref{SSFschw} for the regularized temporal and radial components of the self force (in units of $q$) in the Schwarzschild case ($w=0$) and the numerical values taken from Tables II and III of Ref. \cite{Warburton:2010eq}.
The 2nd and 3rd column display the values obtained by our analytical expressions, whereas the last two columns show the difference with the above mentioned numerical results.
}
\begin{ruledtabular}
\begin{tabular}{ccccc}
$y$ & $F_t^{0\,\rm schw}$ & $F_r^{0\,\rm schw}$ & $\Delta F_t^{0\,\rm schw}$ & $\Delta F_r^{0\,\rm schw}$ \cr
\hline
1/6& $3.088678309\times10^{-4}$& $2.069192430\times10^{-4}$& $5.20\times10^{-5}$& $-3.92\times10^{-5}$\cr 
1/7& $1.621300383\times10^{-4}$& $9.086682872\times10^{-5}$& $1.46\times10^{-5}$& $-1.24\times10^{-5}$\cr 
1/8& $9.278324817\times10^{-5}$& $4.535558187\times10^{-5}$& $4.94\times10^{-6}$& $-4.53\times10^{-6}$\cr 
1/10& $3.667967766\times10^{-5}$& $1.462629728\times10^{-5}$& $8.23\times10^{-7}$& $-8.42\times10^{-7}$\cr 
1/14& $9.180183375\times10^{-6}$& $2.785864322\times10^{-6}$& $5.66\times10^{-8}$& $-6.58\times10^{-8}$\cr 
1/20& $2.148236416\times10^{-6}$& $4.981418996\times10^{-7}$& $3.36\times10^{-9}$& $-4.35\times10^{-9}$\cr
1/30& $4.175425467\times10^{-7}$& $7.191466709\times10^{-8}$& $1.36\times10^{-10}$& $-1.96\times10^{-10}$\cr 
1/50& $5.359926673\times10^{-8}$& $6.350626477\times10^{-9}$& $2.40\times10^{-12}$& $-3.93\times10^{-12}$\cr 
1/70& $1.391199738\times10^{-8}$& $1.284814550\times10^{-9}$& $1.67\times10^{-13}$& $-3.15\times10^{-13}$\cr 
1/100& $3.335029050\times10^{-9}$& $2.356682550\times10^{-10}$& $9.90\times10^{-15}$& $-6.83\times10^{-14}$\cr
\end{tabular}
\end{ruledtabular}
\label{tab:x}
\end{table}


\begin{figure*}
\centering
\includegraphics[scale=0.4]{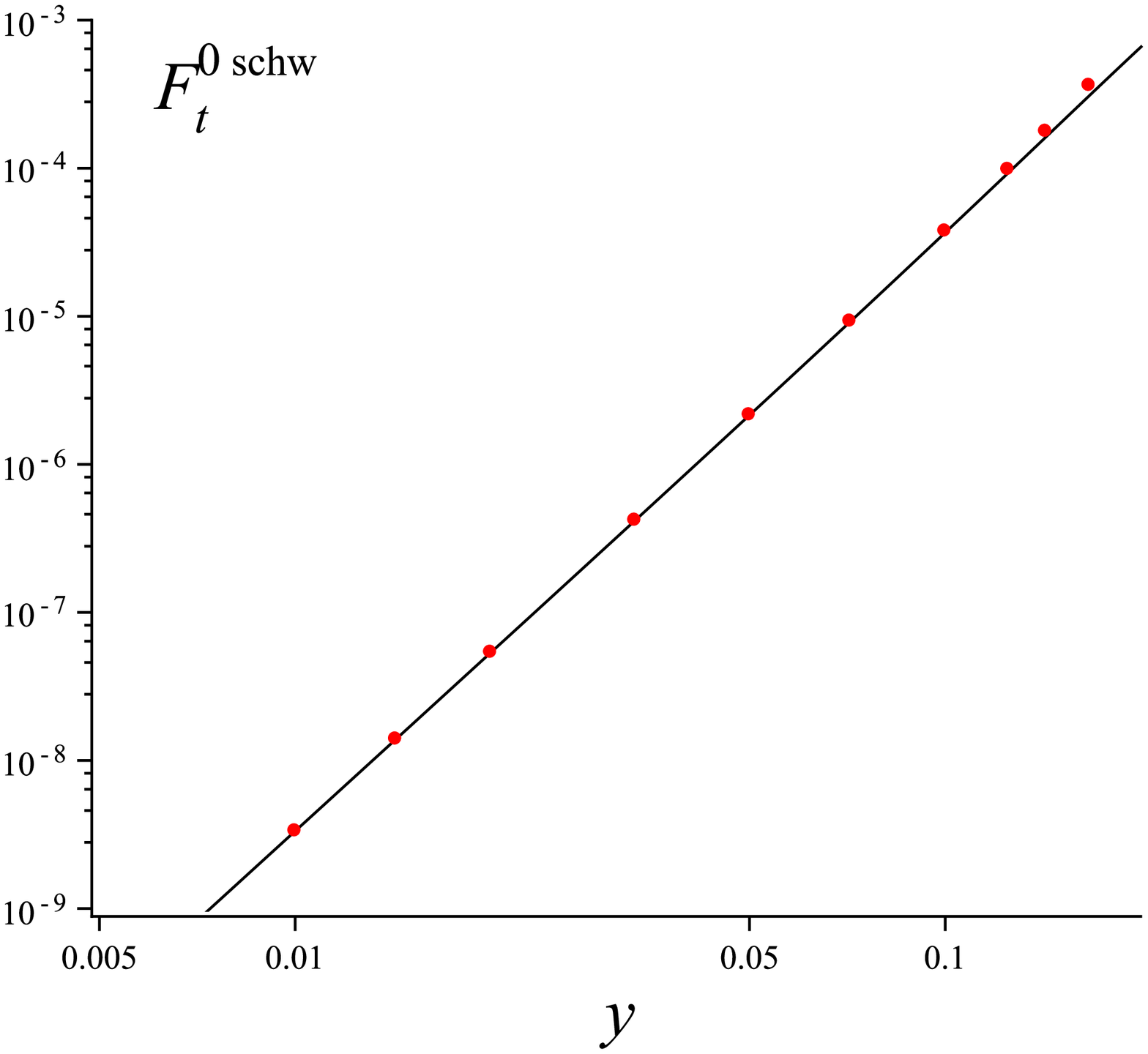}
\includegraphics[scale=0.4]{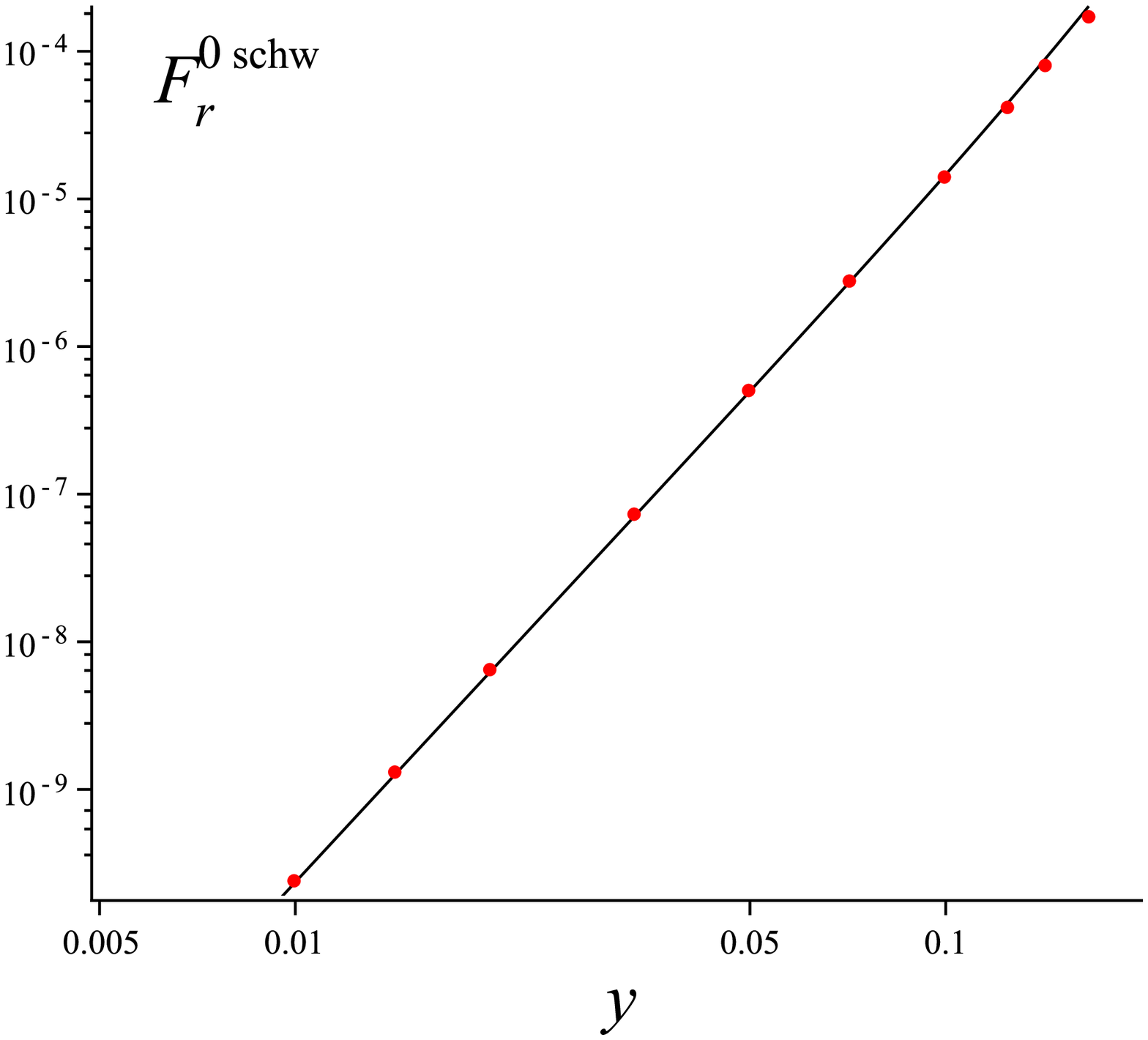}
\caption{
Comparison of numerical data from Ref. \cite{Warburton:2010eq} for the regularized temporal and radial components of the self force (in units of $q$) in the Schwarzschild case ($w=0$) with the behavior of the corresponding analytical expressions \eqref{SSFschw}.
}
\label{fig:2}
\end{figure*}

\end{widetext}

\section{Scalar radiation}

Let us compute the amount of scalar radiation either flowing into the hole or transmitted at spatial infinity. We need to construct the solution to the non-homogeneous wave equation \eqref{radialeq} which satisfies purely ingoing-wave boundary conditions at the black hole horizon and purely outgoing-wave boundary
conditions at infinity.
This is accomplished by using the two kinds of solutions $R^{H,\infty}_{lm\omega}$ to the corresponding homogeneous equation with asymptotic behavior \cite{Teukolsky:1973ha,Press:1973zz,Teukolsky:1974yv}
\begin{eqnarray}
\label{asympsol}
R^H_{lm\omega}&\to&\left\{
\begin{array}{ll}
B^{\rm trans}e^{-i\omega r_*}\,,& r\to r_+\\[3ex]
B^{\rm ref}\displaystyle\frac{e^{i\omega r_*}}{r}+B^{\rm inc}\displaystyle\frac{e^{-i\omega r_*}}{r}\,,& r\to\infty\\
\end{array}
\right.
\,,\nonumber\\[3ex]
R^\infty_{lm\omega}&\to&\left\{
\begin{array}{ll}
C^{\rm up}e^{i\omega r_*}+C^{\rm ref}e^{-i\omega r_*}\,,& r\to r_+\\[3ex]
C^{\rm trans}\displaystyle\frac{e^{i\omega r_*}}{r}\,,& r\to\infty\\
\end{array}
\right.
\,,
\end{eqnarray}
where $r_*$ is the tortoise-like coordinate defined by ${dr_*}/{dr}={r^2}/{\Delta}$, i.e.,
\beq
\label{rstardef}
r_*=r+\frac{2Mr_+}{r_+-r_-}\ln\frac{r-r_+}{2M}-\frac{2Mr_-}{r_+-r_-}\ln\frac{r-r_-}{2M}\,.
\eeq
The final solution is given by \cite{Sasaki:2003xr} 
\beq
R_{lm\omega}(r)=Z^H_{lm\omega}(r)R^\infty_{lm\omega}(r)+Z^\infty_{lm\omega}(r)R^H_{lm\omega}(r)\,,
\eeq
where
\begin{eqnarray}
Z^H_{lm\omega}(r)&=&\frac{-1}{W_{lm\omega}}\int_{r_+}^{r}R^H_{lm\omega}(r')\Delta(r')S_{lm\omega}\delta(r' - r_0)dr'\,,\nonumber\\
Z^\infty_{lm\omega}(r)&=&\frac{-1}{W_{lm\omega}}\int_{r}^{\infty}R^\infty_{lm\omega}(r')\Delta(r')S_{lm\omega}\delta(r' - r_0)dr'\,,\nonumber\\
\end{eqnarray}
and $W_{lm\omega}=2i\omega C^{\rm trans}B^{\rm inc}$ is the constant Wronskian.
The asymptotic behaviors of $R_{lm\omega}$ at the horizon and at infinity are then 
\begin{eqnarray}
R_{lm\omega}(r\to r_+)&\to& B^{\rm trans}Z^\infty_{lm\omega}(r_+)e^{-i\omega r_*}\,,\nonumber\\
R_{lm\omega}(r\to \infty)&\to& C^{\rm trans}Z^H_{lm\omega}(\infty)\frac{e^{i\omega r_*}}{r}\,,
\end{eqnarray}
respectively, so that one can define the amplitudes
\beq
{\mathcal Z}^H_{lm\omega}=B^{\rm trans}Z^\infty_{lm\omega}(r_+)\,,\qquad
{\mathcal Z}^\infty_{lm\omega}=C^{\rm trans}Z^H_{lm\omega}(\infty)\,.
\eeq
Explicitly we find
\begin{eqnarray}
\label{tildezeta}
{\mathcal Z}^H_{lm\omega}&=&2\pi \frac{B^{\rm trans}}{W_{lm\omega}}\,r_0q_{lm\omega}R^\infty_{lm\omega}(r_0)\delta(\omega-m\Omega)\nonumber\\
&\equiv&2\pi \tilde{\mathcal Z}^H_{lm}\delta(\omega-m\Omega)\,,
\nonumber\\
{\mathcal Z}^\infty_{lm\omega}&=&2\pi \frac{C^{\rm trans}}{W_{lm\omega}}\,r_0q_{lm\omega}R^H_{lm\omega}(r_0)\delta(\omega-m\Omega)\nonumber\\
&\equiv&2\pi \tilde{\mathcal Z}^\infty_{lm}\delta(\omega-m\Omega)\,.
\end{eqnarray}
The energy flux at infinity is thus given by \cite{Teukolsky:1973ha,Press:1973zz,Teukolsky:1974yv}
\beq
\label{infflux}
\frac{dE^\infty}{dt}=\sum_{lm}\frac{\omega^2}{4\pi}|\tilde{\mathcal Z}^\infty_{lm}|^2\,,
\eeq
while the energy flux at the event horizon is
\beq
\label{horflux}
\frac{dE^H}{dt}=\sum_{lm}\frac{M\omega^2r_+}{2\pi}|\tilde{\mathcal Z}^H_{lm}|^2\,,
\eeq
with $\tilde{\mathcal Z}^{H,\infty}_{lm}$ defined in Eq. \eqref{tildezeta} and $\omega=m\Omega$.

For the computation of the amplitudes and the transmission coefficients we have used the MST ingoing ad upgoing solutions, which satisfy the proper boundary conditions at the horizon and at infinity for any given value of $l$, i.e.,  $R^{H}_{lm\omega}(r)=R^{\rm in(MST)}_{lm\omega}(r)$ and $R^{\infty}_{lm\omega}(r)=R^{\rm up(MST)}_{lm\omega}(r)$.
The corresponding transmission coefficients are given by \cite{Sasaki:2003xr}
\begin{eqnarray}
B^{\rm trans}&=&e^{i\frac{\kappa}{2}(\epsilon+\tau)\left(1+\frac{2\ln\kappa}{1+\kappa}\right)}\sum_{n=-\infty}^{\infty}a_n\,,
\nonumber\\
C^{\rm trans}&=&\omega^{-1}e^{i\epsilon\left(\ln\epsilon-\frac{1-\kappa}2\right)}A_-^\nu\,,
\end{eqnarray}
where
\beq
A_-^\nu=2^{-1+i\epsilon}e^{-\frac{\pi}{2}i(\nu+1)}e^{-\frac{\pi}{2}\epsilon}\sum_{n=-\infty}^{\infty}(-1)^n\frac{(\nu+1-i\epsilon)_n}{(\nu+1+i\epsilon)_n}\,a_n\,.
\eeq
The definitions of the various quantities $\epsilon, \tau, \nu, a_n$ are given in Appendix \ref{app:MST} for convenience.

We find (in units of $q$)

\begin{widetext}

\begin{eqnarray}
\frac{dE^\infty}{dt}&=&\left(\frac{dE^\infty}{dt}\right)_N\left\{
1+\left(-2+\frac{2}{3}w\right)y+2\pi y^{3/2}
+\left(-10+\frac{13}{3}w-\frac13w^2\right)y^2
+\left(\frac{12}{5}+\frac43w\right)\pi y^{5/2}\right.\nonumber\\
&&
+\left[\frac{1331}{75}-\frac{2203}{225}w+\frac49w^2+\frac{28}{81}w^3+\frac43\pi^2\right.\nonumber\\
&&
\left.
+\left(-\frac{76}{15}+\frac{16}{15}w\right)\gamma+\left(-\frac{76}{15}+\frac{16}{15}w\right)\ln(2)+\left(-\frac{38}{15}+\frac{8}{15}w\right)\ln(y)\right]y^3\nonumber\\
&&\left.
+\left(-\frac{521}{14}+\frac{86}{5}w-\frac23w^2\right)\pi y^{7/2}
+O(y^4)
\right\}\,,
\nonumber\\
\frac{dE^H}{dt}&=&\left(\frac{dE^\infty}{dt}\right)_N 2(1-w)(1+\sqrt{1-w})\left\{
y^3+\left(2-\frac43w\right)y^4+\left(3-\frac73w+2w^2\right)y^5
\right.\nonumber\\
&& 
+\left[\frac{1231}{75}-\frac{793}{225}w+\frac{32}{9}w^2-\frac{260}{81}w^3+\frac{1}{1-w}
+\left(-\frac{38}{15}+\frac{8}{15}w\right)\ln(1-w)
+\frac1{(1-w)^{1/2}}\left(\frac43-\frac23w\right)\pi^2\right.\nonumber\\
&&\left.\left.
-\frac83\gamma+\left(-\frac{116}{15}+\frac{16}{15}w\right)\ln(2) +\left(-\frac{32}{5}+\frac{16}{15}w\right)\ln(y)\right]y^6
\right.\nonumber\\
&&\left.
+\left[
\frac{1769}{225}-\frac{46}{15}w+\frac{907}{675}w^2-\frac{473}{81}w^3+\frac{1309}{243}w^4+\frac{1}{1-w}\left(4-\frac{10}{3}w\right)
+\left(-\frac{76}{15}+\frac{40}{9}w-\frac{32}{45}w^2\right)\ln(1-w)\right.\right.\nonumber\\
&&\left.\left.
+\frac1{(1-w)^{1/2}}\left(\frac83-\frac{28}{9}w+\frac89w^2\right)\pi^2
+\left(\frac{64}{15}+\frac89w\right)\gamma
+\left(-\frac{88}{15}+\frac{88}{9}w-\frac{64}{45}w^2\right)\ln(2)\right.\right.\nonumber\\
&&\left.\left.
+\left(-8+\frac{28}{3}w-\frac{64}{45}w^2\right)\ln(y)
\right]y^7
\right.\nonumber\\
&&\left.
+\left(-\frac{56}{45}+\frac{16}{45}w\right)\pi y^{15/2}
\right.\nonumber\\
&&\left.
+\left[
-\frac{208762}{7875}+\frac{10034656}{165375}w-\frac{4518949}{165375}w^2+\frac{1162}{675}w^3+10w^4-\frac{28}{3}w^5
+\frac{1}{1-w}\left(9-\frac{44}{3}w+\frac{25}{3}w^2\right)\right.\right.\nonumber\\
&&\left.\left.
+\left(-\frac{1818}{175}+\frac{17086}{1575}w-\frac{10804}{1575}w^2+\frac{16}{15}w^3\right)\ln(1-w)
+\frac1{(1-w)^{1/2}}\left(\frac{36}{5}-\frac{446}{45}w+\frac{262}{45}w^2-\frac43w^3\right)\pi^2\right.\right.\nonumber\\
&&\left.\left.
+\left(\frac{608}{105}-\frac{344}{45}w-\frac89w^2\right)\gamma
+\left(-\frac{1124}{75}+\frac{22132}{1575}w-\frac{23008}{1575}w^2+\frac{32}{15}w^3\right)\ln(2)\right.\right.\nonumber\\
&&\left.\left.
+\left(-\frac{9388}{525}+\frac{3128}{175}w-\frac{7436}{525}w^2+\frac{32}{15}w^3\right)\ln(y)
\right]y^8
\right.\nonumber\\
&&\left.
+\left(\frac{88}{225}+\frac{196}{675}w-\frac{16}{135}w^2\right)\pi y^{17/2}
+O(y^9)
\right\}\,,
\end{eqnarray}
where
\beq
\left(\frac{dE^\infty}{dt}\right)_N=\frac13y^4\,,
\eeq
in terms of the gauge-invariant variable $y$ (see Eq. \eqref{uvsy}).
Note that the flux at infinity is computed up to the 3.5PN order, i.e., at $O(y^{7/2})$ included (see Appendix \ref{app:fluxes}). 
The leading contribution to the flux on the horizon, instead, enters at 3PN order beyond the lowest order, and is computed through $O(y^{17/2})$.  

In the Schwarzschild case ($w\to0$) the previous expressions reduce to
\begin{eqnarray}
\label{schwfluxes}
\frac{dE^\infty}{dt}&=&\left(\frac{dE^\infty}{dt}\right)_N\left[
1-2y+2\pi y^{3/2}-10y^2+\frac{12}{5}\pi y^{5/2}+\left(\frac{1331}{75}+\frac43\pi^2-\frac{76}{15}\gamma-\frac{76}{15}\ln(2)-\frac{38}{15}\ln(y)\right)y^3\right.\nonumber\\
&&\left.
-\frac{521}{14}\pi y^{7/2}
+O(y^4)
\right]\,,
\nonumber\\
\frac{dE^H}{dt}&=&\left(\frac{dE^\infty}{dt}\right)_N 4y^3\left[
1+2y+3y^2+\left(\frac{1306}{75}+\frac43\pi^2-\frac{8}{3}\gamma-\frac{116}{15}\ln(2)-\frac{32}{5}\ln(y)\right)y^3\right.\nonumber\\
&&\left.
+\left(\frac{2669}{225}+\frac83\pi^2+\frac{64}{15}\gamma-\frac{88}{15}\ln(2)-8\ln(y)\right)y^4
-\frac{56}{45}\pi y^{9/2}\right.\nonumber\\
&&\left.
+\left(-\frac{137887}{7875}+\frac{36}{5}\pi^2+\frac{608}{105}\gamma-\frac{1124}{75}\ln(2)-\frac{9388}{525}\ln(y)\right)y^5
+\frac{88}{225}\pi y^{11/2}
+O(y^6)
\right]\,,
\end{eqnarray}
whereas in the extreme RN case ($w\to1$) we have
\begin{eqnarray}
\frac{dE^\infty}{dt}&=&\left(\frac{dE^\infty}{dt}\right)_N\left[
1-\frac43y+2\pi y^{3/2}-6y^2+\frac{56}{15}\pi y^{5/2}+\left(\frac{3542}{405}+\frac43\pi^2-4\gamma-4\ln(2)-2\ln(y)\right)y^3\right.\nonumber\\
&&\left.
-\frac{4343}{210}\pi y^{7/2}
+O(y^4)
\right]\,,
\nonumber\\
\frac{dE^H}{dt}&=&\left(\frac{dE^\infty}{dt}\right)_N 2y^6\left[
1+\frac23y+\frac83y^2
+O(y^3)
\right]\,.
\end{eqnarray}
Therefore, when the black hole is extremely charged the horizon-absorbed flux starts 3PN orders more beyond with respect to the non-extreme case. 

Finally, we note that the angular momentum fluxes can be easily calculated through
\beq
\frac{dJ^{H,\infty}}{dt}=y^{-3/2}\frac{dE^{H,\infty}}{dt}\,.
\eeq

  
\begin{table}
\centering
\caption{
Comparison between the analytical expressions \eqref{schwfluxes} for the energy fluxes at infinity and on the horizon (in units of $q$) in the Schwarzschild case ($w=0$) and the numerical values taken from Table I of Ref. \cite{Warburton:2010eq}.
The 2nd and 3rd column display the values of the flux radiated to infinity $\dot E^\infty$ and down to the horizon $\dot E^H$ (divided by the total flux $\dot E^{\rm tot}=\dot E^\infty+\dot E^H$), respectively, obtained by our analytical expressions, whereas the last two columns show the difference with the above mentioned numerical results. 
}
\begin{ruledtabular}
\begin{tabular}{ccccc}
$y$ & $\dot E^\infty$ & $\dot E^H/\dot E^{\rm tot}$ & $\Delta\dot E^\infty$ & $\Delta(\dot E^H/\dot E^{\rm tot})$ \cr
\hline
 1/6  & $2.096811267\times10^{-4}$ & $0.03565284588$ & $4.55\times10^{-5}$ & $-4.85\times10^{-3}$\cr
 1/8  & $7.246350703\times10^{-5}$ & $0.01200009767$ & $4.79\times10^{-6}$ & $-6.00\times10^{-4}$ \cr
 1/10  & $3.052889221\times10^{-5}$ & $5.536219295\times10^{-3}$ & $8.48\times10^{-7}$ & $-1.36\times10^{-4}$  \cr
 1/20   &$1.979588741\times10^{-6}$ & $5.854584422\times10^{-4}$ & $4.08\times10^{-9}$ & $1.45\times10^{-5}$  \cr
 1/40   & $1.262462352\times10^{-7}$ & $6.785017170\times10^{-5}$ & $-1.95\times10^{-11}$ & $3.21\times10^{-5}$  \cr
\end{tabular}
\end{ruledtabular}
\label{tab:2}
\end{table}

\end{widetext}

\section{Concluding remarks}

We have analyzed scalar self-force effects on a scalar charge moving along a circular orbit around a Reissner-Nordstr\"om black hole. 
The scalar wave equation is separated by using standard spherical harmonics (available here because of the underlying spherical symmetry of the background) and the field is decomposed into frequency modes. The associated radial equation is solved perturbatively in a PN framework by using the Green function method.
The scalar field as well as the components of the self-force are then regularized at the particle's position by subtracting the divergent term mode by mode, summing the infinite series up to a certain PN order.
The MST approach has also been adopted for computing a number of radiative multipoles (up to $l=4$), so that our final result is accurate up to the 7.5PN order, i.e., up to the order $O(y^{15/2})$ included in terms of the dimensionless gauge-invariant frequency variable $y=(M\Omega)^{2/3}$.
Since the scalar charge interacts only gravitationally with the background field, the coupling with the black hole electromagnetic charge is quadratic.
The two limiting cases of a Schwarzschild black hole (which was missing in the literature and represents by itself an interesting byproduct of our work) and of an extreme Reissner-Nordstr\"om black hole are discussed explicitly. The comparison of the analytically computed regularized field and self-force components with existing results in the Schwarzschild case \cite{Detweiler:2002gi,Warburton:2010eq} has shown a good agreement, the difference between analytically and numerically produced values ranging from $\sim 10^{-14}$ in the weak-field region to $\sim 10^{-4}$ in the strong-field region.

We have also evaluated the radiation fluxes both at infinity and on the outer horizon up to $O(y^{7/2})$ and $O(y^{17/2})$ included, respectively.
We have found that, when the black hole is extremely charged, the horizon-absorbed flux starts 3PN orders more beyond than the non-extreme case.
The comparison with previous numerical results in the Schwarzschild case \cite{Warburton:2010eq} has shown again a good agreement in both weak-field and strong-field regimes.

\section*{Acknowledgements}
BD thanks ICRANet for partial support.
GC is  grateful to the Brazilian agency CAPES for the financial support in form of a doctoral scholarship BEX-$15114/13-9$.
All authors thank Prof. T. Damour for useful comments.

\appendix

\begin{widetext}

\section{Homogeneous solutions to the scalar wave equation}

\subsection{PN solutions}

PN solutions have the form \eqref{PNsols}, i.e.,
\begin{eqnarray}
R_{\rm in(PN)}^{lm\omega}(r) &=& r^l [1 + A_2^{lm\omega}(r)\eta^2 + A_4^{lm\omega}(r)\eta^4 + A_6^{lm\omega}(r)\eta^6 + A_8^{lm\omega}(r)\eta^8 + \ldots]\,,\nonumber\\
R_{\rm up(PN)}^{lm\omega}(r) &=& R_{\rm in(PN)}^{-l-1m\omega}(r)\,. 
\end{eqnarray}
The first coefficients are given by
\begin{eqnarray}
A_2^{lm\omega}(r) &=&
-\frac{Ml}{r}-\frac{\omega^2r^2}{2(2l+3)}
\,,\nonumber\\
A_4^{lm\omega}(r) &=&
\frac{M^2}{r^2}\frac{l(l-1)(2l-1-\kappa^2)}{2(2l-1)}
+M\omega^2r\frac{l^2-5l-10}{2(2l+3)(l+1)}
+\frac{\omega^4r^4}{8(2l+3)(2l+5)}
\,,\nonumber\\
A_6^{lm\omega}(r) &=&
-\frac{M^3}{r^3}\frac{l(l-1)(l-2)(2l-1-3\kappa^2)}{6(2l-1)}
-2M^2\omega^2\frac{[3(2l-1)(2l+3)+(3l^2+3l-2)\kappa^2][(2l+1)\ln(r/R)-1]}{(2l-1)(2l+3)(2l+1)^2}\nonumber\\
&&
-M\omega^4r^3\frac{3l^3-27l^2-142l-136}{24(l+1)(l+2)(2l+3)(2l+5)}
-\frac{\omega^6r^6}{48(2l+3)(2l+5)(2l+7)}
\,,\nonumber\\
A_8^{lm\omega}(r) &=&
\frac{M^4}{r^4}\frac{l(l-1)(l-2)(l-3)}{24(2l-1)(2l-3)}[(2l-1)(2l-3)-3\kappa^2(4l-6-\kappa^2)]\nonumber\\
&&
+\frac{M^3\omega^2}{r}\left\{
-\frac{4l^6-32l^5-99l^4-241l^3-436l^2-276l-36}{6l(2l+3)(2l+1)^2}
+\frac{48l^5+348l^4+540l^3+126l^2-120l-36}{6l(2l+3)(2l-1)(2l+1)^2}\kappa^2\right.\nonumber\\
&&\left.
+\left[
\frac{6l}{2l+1}
+\frac{2l(3l^2+3l-2)}{(2l+3)(2l-1)(2l+1)}\kappa^2
\right]\ln(r/R)
\right\}\nonumber\\
&&
+M^2\omega^2r^2\left\{
-\frac{24l^7+156l^6-1766l^5-13267l^4-29512l^3-23465l^2-2058l+2784}{48(l+1)(2l+3)^2(2l+5)(2l+1)^2(l+2)}\right.\nonumber\\
&&
+\frac{16l^6+80l^5-440l^4-2432l^3-2803l^2+11l+784}{16(2l-1)(2l+3)^3(2l+5)(2l+1)^2}\kappa^2\nonumber\\
&&\left.
+\left[
\frac{3}{(2l+3)(2l+1)}
+\frac{3l^2+3l-2}{(2l-1)(2l+3)^2(2l+1)}\kappa^2
\right]\ln(r/R)
\right\}
\nonumber\\
&&
+M\omega^6r^5\frac{5l^4-60l^3-625l^2-1548l-1108}{240(l+3)(l+2)(2l+7)(2l+5)(2l+3)(l+1)}
+\frac{\omega^8r^8}{384(2l+9)(2l+7)(2l+5)(2l+3)}
\,, 
\end{eqnarray}
where $R$ is a length scale.
This solution, which we need however to compute the sum over all multipoles, becomes immediately inadequate, and one should use the MST technology. In fact, the coefficient $A_4$ of the \lq\lq up'' solution (obtained from $A_4^{lm\omega}(r)$ with $l\to-l-1$) diverges for $l=0$; similarly, higher order coefficients diverge for $l=1,2,\ldots$ etc.

\subsection{MST solutions}
\label{app:MST}

The MST technique \cite{Mano:1996mf,Mano:1996vt} allows to find homogeneous solution to the radial equation which satisfy retarded boundary conditions at the horizon ($R_{\rm in(MST)}^{lm\omega}(r)$) and radiative boundary conditions at infinity ($R_{\rm up(MST)}^{lm\omega}(r)$).

The ingoing solution can be formally written as a convergent (at any finite value of $r$) series of hypergeometric functions
\beq
\label{MSTin}
R_{\rm in(MST)}^{lm\omega}(x) =  C_{\rm(in)}(x)\sum_{n=-\infty}^{\infty}a_{n}
F(n+\nu+1-i\tau,-n-\nu-i\tau,1-i\epsilon-i\tau;x)\,,
\eeq
with 
\beq
C_{\rm(in)}(x)=e^{i\epsilon\kappa x}(-x)^{-i(\epsilon+\tau)/2}(1-x)^{i(\epsilon-\tau)/2}\,,
\eeq
where the new variable $x=(r_+-r)/2M\kappa$ has been introduced and  
\beq
\epsilon= 2M\omega\,,\quad
\tau=\frac{1}{2}\frac{\epsilon(\kappa^2+1)}{\kappa}\,.
\eeq 
The hypergeometric functions above are better evaluated by using the standard identity 
\beq
\label{hyp_y}
F(a,b;c; x)=y^a \frac{\Gamma (c) \Gamma (b-a)}{\Gamma (b) \Gamma (c-a)}F(a,c-b,a-b+1;y)+ y^b \frac{\Gamma (c) \Gamma (a-b)}{\Gamma (a) \Gamma (c-b)}F(b,c-a,b-a+1;y)\,,
\eeq
involving the \lq\lq small'' variable $y=1/(1-x)$.
Note that the overall factor $\Gamma (c)$ does not depend on $n$, so that it can be factored out.

The expansion coefficients $a_{n}$ satisfy the following three-term recurrence relation
\beq
\alpha_n^{\nu}a_{n+1} + \beta_n^{\nu}a_n+\gamma_n^{\nu}a_{n-1} =0\,,
\eeq
where
\begin{eqnarray}
\alpha_n^{\nu}&=&
\frac{i\epsilon\kappa(n+\nu+1+i\epsilon)(n+\nu+1-i\epsilon)(n+\nu+1+i\tau)}{(n+\nu+1)(2n+2\nu+3)}
\,,\nonumber\\
\beta_n^{\nu}&=&
-l(l+1)+(n+\nu)(n+\nu+1)+\epsilon\kappa\tau+\epsilon^2+\frac{\epsilon^3\kappa\tau}{(n+\nu)(n+\nu+1)}
\,,\nonumber\\
\gamma_n^{\nu}&=&
-\frac{i\epsilon\kappa(n+\nu+i\epsilon)(n+\nu-i\tau)(n+\nu-i\epsilon)}{(n+\nu)(2n+2\nu-1)}
\,.
\end{eqnarray}
Once the recurrence system has been solved for $n = 1\ldots N$ and $n = -N\ldots-1$ for a given $N$ such that $a_N=0=a_{-N}$, the case $n = 0$ with $a_0=1$ becomes a compatibility condition which yields the parameter 
\beq
\nu=l+\sum _{k=1}^{\infty}\nu_k \epsilon^{2k}\,.
\eeq
The solution of the recurrence system is rather involved (even in this relatively simple case). 
The structure of the expansion coefficients 
\beq
\label{ansymb}
a_{n}=\sum _{k=i}^{j}c_{nk} \epsilon^k
\eeq
is summarized in Table \ref{tab:3} for $l=1$ and $N=15$, as an example.

  
\begin{table}[h]
\centering
\caption{The structure of the expansion coefficients \eqref{ansymb} of the recurrence relation is shown for $l=1$ and $N=15$, so that $a_{-15}=0=a_{15}$ and $a_0=1$.
}
\begin{ruledtabular}
\begin{tabular}{cccccccccccccccccccccccccccccc}
$n$ & $-14$ & $-13$ & $-12$ & $-11$ & $-10$ & $-9$ & $-8$ & $-7$ & $-6$ & $-5$ & $-4$ & $-3$ & $-2$ & $-1$ & $0$ & $1$ & $2$ & $3$ & $4$ & $5$ & $6$ & $7$ & $8$ & $9$ & $10$ & $11$ & $12$ & $13$ & $14$ \cr
$i$ & $13$ & $12$ & $11$ & $10$ & $9$ & $8$ & $7$ & $6$ & $5$ & $4$ & $3$ & $2$ & $3$ & $1$ & - & $1$ & $2$ & $3$ & $4$ & $5$ & $6$ & $7$ & $8$ & $9$ & $10$ & $11$ & $12$ & $13$ & $14$ \cr
$j$ & $14$ & $14$ & $14$ & $14$ & $14$ & $14$ & $14$ & $14$ & $14$ & $14$ & $14$ & $14$ & $14$ & $14$ & - & $14$ & $14$ & $14$ & $14$ & $14$ & $14$ & $14$ & $14$ & $14$ & $14$ & $14$ & $14$ & $14$ & $14$ \cr
\end{tabular}
\end{ruledtabular}
\label{tab:3}
\end{table}

The upgoing solution can be formally written as a convergent (at spatial infinity) series of irregular confluent hypergeometric functions with the same
series coefficients
\beq
\label{MSTup}
R_{\rm up(MST)}^{lm\omega}(z) =  C_{\rm(up)}(z)\sum_{n=-\infty}^{\infty}a_{n}
\frac{(\nu+1-i\epsilon)_n}{(\nu+1+i\epsilon)_n}\,  (2iz)^n\Psi[n+\nu+1-i\epsilon,2n+2\nu+2; -2i z]\,,
\eeq
with
\beq
C_{\rm (up)}(z)=(2z)^\nu e^{-\pi \epsilon} e^{-i\pi (\nu+1)}e^{i z} \left(1-\frac{\epsilon \kappa}{z}\right)^{-i(\epsilon+\tau)/2}\,,
\eeq
where the new variable $z=\omega(r-r_-)=\epsilon\kappa(1-x)$ has been introduced and $(A)_n=\Gamma (A+n)/\Gamma(A)$ is the Pochammer symbol.
The irregular confluent hypergeometric functions above can be conveniently split into two pieces by using the identity
\beq
\Psi(a,b;\zeta)=\frac{\Gamma (1-b)}{\Gamma(a-b+1)}F(a,b;\zeta)+\zeta^{1-b}\frac{\Gamma(b-1)}{\Gamma(a)}F(a-b+1,2-b;\zeta)\,.
\eeq

For instance, for $l=1$ we get
\begin{eqnarray}
R_{\rm in(MST)}^{l=1}(r) &=&
\frac{r}{M\kappa}
-\frac{1}{\kappa}\left(1+\frac{\omega^2r^3}{10M}\right)\eta^2
+\frac{i\omega r}{\kappa^2}\left[1+2\kappa+2\kappa^2-(1+\kappa)^2\gamma\right]\eta^3
-\frac{\omega^2 r^2}{\kappa}\left(\frac{7}{10}-\frac{\omega^2r^3}{280M}\right)\eta^4\nonumber\\
&&
-\frac{iM\omega}{\kappa^2}\left(1+\frac{\omega^2r^3}{10M}\right)\left[1+2\kappa+2\kappa^2-(1+\kappa)^2\gamma\right]\eta^5\nonumber\\
&&
-\left\{
\left[\frac{16875+120375\kappa^2+87075\kappa^4+23790\kappa^6+1616\kappa^8}{75(15+4\kappa^2)^2}
+\left[1+2\kappa+2\kappa^2-(1+\kappa)^2\gamma\right]^2\right.\right.\nonumber\\
&&\left.\left.
-\frac16(1+6\kappa^2+\kappa^4)\pi^2
-\frac{4}{15}\kappa^2(15+4\kappa^2)\ln\left(\frac{2M\kappa\eta^2}{r}\right)
\right]\frac{M\omega^2r}{2\kappa^3}
-\frac{\omega^4 r^4}{\kappa}\left(\frac{151}{2520}-\frac{\omega^2r^3}{15120M}\right)
\right\}\eta^6\nonumber\\
&&
+O(\eta^7)
\,, \nonumber\\
R_{\rm up(MST)}^{l=1}(r) &=&
-\frac{i}{2\omega^2r^2}
-\frac{i}{4}\left(1+\frac{4M}{\omega^2r^3}\right)\eta^2
+\frac{M}{\omega r^2}\left(1+\frac{\omega^2r^3}{6M}-\gamma+i\pi\right)\eta^3
-i\left(\frac{M}{r}-\frac{\omega^2r^2}{16}+\frac{3(5+\kappa^2)M^2}{10\omega^2r^4}\right)\eta^4\nonumber\\
&&
+\left[\left(\frac13-\frac{\gamma}{2}+\frac{i\pi}{2}\right)M\omega
+\left(1-\gamma+i\pi\right)\frac{2M^2}{\omega r^3}
-\frac{\omega^3r^3}{60}
\right]\eta^5\nonumber\\
&&
+\left\{
-\frac{2iM^3}{5\omega^2r^5}(5+3\kappa^2)
+\frac{iM^2}{15r^2}\left[
-(15+4\kappa^2)\ln(2\omega r\eta)
+15(\gamma-i\pi)^2-(45+4\kappa^2)\gamma+30i\pi-\frac{5}{2}\pi^2\right.\right.\nonumber\\
&&\left.\left.
+\frac{101250+67800\kappa^2+15435\kappa^4+848\kappa^6}{10(15+4\kappa^2)^2}\right]
+\frac{iM\omega^2r}{3}\left(2\ln(2\omega r\eta)+\gamma-\frac{61}{24}\right)
-\frac{i\omega^4r^4}{288}
\right\}\eta^6\nonumber\\
&&
+O(\eta^7)
\,,
\end{eqnarray}
having rescaled the \lq\lq in'' solution by the constant factor $\Gamma (c)$, with 
\beq
\nu=1
-\left(\frac12+\frac{2}{15}\kappa^2\right)\epsilon^2
-\frac{496125+680400\kappa^2+135990\kappa^4+12688\kappa^6}{189000(15+4\kappa^2)}\epsilon^4
+O(\epsilon^6)\,.
\eeq

\section{Energy fluxes}
\label{app:fluxes}

Using the notation of Ref. \cite{Sasaki:2003xr}, the energy fluxes \eqref{infflux} and \eqref{horflux} can be written as 
\begin{eqnarray}
\frac{dE^\infty}{dt}&=&\left(\frac{dE^\infty}{dt}\right)_N\sum_{l=1}^\infty\sum_{m=-l}^l\eta^\infty_{lm}\,,\nonumber\\
\frac{dE^H}{dt}&=&\left(\frac{dE^\infty}{dt}\right)_N2(1-w)(1+\sqrt{1-w})y^3\sum_{l=1}^\infty\sum_{m=-l}^l\eta^H_{lm}\,,
\end{eqnarray}
with $\eta^{H,\infty}_{l-m}=\eta^{H,\infty}_{lm}$.
For small values of the dimensionless angular velocity variable $y$ the expansion coefficients behave as $\eta^\infty_{lm}\sim y^{l-1}$ and $\eta^H_{lm}\sim y^{2(l-1)}$, for fixed values of $m$.
Therefore, since we have used the MST solutions up to $l=4$, our calculations of the flux at infinity and on the horizon are accurate up to the order $O(y^{7/2})$  and $O(y^{15/2})$ beyond the lowest order, respectively. 
For example, for $l=1$ the first few terms of the expansion are given by
\begin{eqnarray}
\eta^\infty_{11}&=&
\frac12+\left(-\frac{13}{5}+\frac13w\right)y+\pi y^{3/2}+\left(\frac{1123}{350}+\frac{1}{30}w-\frac16w^2\right)y^2+\left(-\frac{26}{5}+\frac23w\right)\pi y^{5/2}\nonumber\\
&&
+\left[
\frac{10958}{945}-\frac{12427}{3150}w+\frac{26}{45}w^2+\frac{14}{81}w^3
+\frac23\pi^2+\left(-\frac{38}{15}+\frac{8}{15}w\right)\gamma
+\left(-\frac{38}{15}+\frac{8}{15}w\right)\ln(2)\right.\nonumber\\
&&\left.
+\left(-\frac{19}{15}+\frac{4}{15}w\right)\ln(y)
\right]y^3
+\left(\frac{1123}{175}+\frac{1}{15}w-\frac13w^2\right)\pi y^{7/2}
+O(y^4)
\,,\nonumber\\
\eta^H_{11}&=&
\frac12+\left(1-\frac23w\right)y+\left(\frac{11}{10}-\frac{23}{30}w+w^2\right)y^3\nonumber\\
&&
+\left[
\frac{971}{150}+\frac{347}{450}w+\frac{44}{45}w^2-\frac{130}{81}w^3
+\frac{1}{2(1-w)}+\left(-\frac{19}{15}+\frac{4}{15}w\right)\ln(1-w)
+\frac1{(1-w)^{1/2}}\left(\frac23-\frac13w\right)\pi^2\right.\nonumber\\
&&\left.
-\frac43\gamma+\left(-\frac{58}{15}+\frac{8}{15}w\right)\ln(2)+\left(-\frac{16}{5}+\frac{8}{15}w\right)\ln(y)
\right]y^4
+O(y^5)
\,.
\end{eqnarray}

\end{widetext}

\end{document}